\documentstyle[graphicx,amsmath,amsfonts]{article}
\bibliography{plain}
\title{Canonical Coset Parameterization
and the Bures Metric of the Three-level Quantum Systems}
\vspace{20mm}
\author{
  S. J. Akhtarshenas
\thanks{E-mail:akhtarshenas@phys.ui.ac.ir}
\\
{\small Department of Physics, University of Isfahan, Isfahan,
Iran } }
\begin{document}
\maketitle
%\vspace{15mm}
%\newpage

\begin{abstract}
An explicit parameterization for the state space of an $n$-level
density matrix is given. The parameterization is based on the
canonical coset decomposition of unitary matrices. We also
compute, explicitly, the Bures metric tensor over the state space
of two- and three-level quantum systems.


\end{abstract}
%\pagebreak
%\vspace{7cm}
\section{Introduction}
In recent years, the Riemannian Bures metric \cite{bures} has
become an interesting subject for the understanding of the
geometry of quantum state space. It is the quantum analog of
Fisher information in classical statistics, i.e. in the subspace
of diagonal matrices it induces the statistical distance
\cite{braun}. The Bures measure is monotone in the sense that it
does not increase under the action of completely positive, trace
preserving maps \cite{petz}. It is, indeed, minimal among all
monotone metrics and its extension to pure state is exactly the
Fubini-Study metric \cite{petz}. The Bures distance between any
two mixed states $\rho_1$ and $\rho_2$ is a function of their
fidelity $F(\rho_1,\rho_2)$ \cite{uhlmann,jozsa}
\begin{equation}\label{fidelity}
\textmd{d}_{B}(\rho_1,\rho_2)=\sqrt{2-2\sqrt{F(\rho_1,\rho_2)}},\qquad
F(\rho_1,\rho_2)=
\left[\textmd{Tr}\left(\sqrt{\sqrt{\rho_1}\rho_2\sqrt{\rho_1}}\right)\right]^2.
\end{equation}
Fidelity allows one to characterize the closeness of the pair of
mixed states $\rho_1$ and $\rho_2$, so, it is an important concept
in quantum mechanics, quantum optics and quantum information
theory. An explicit formula for the infinitesimal Bures distance
between $\rho$ and $\rho+\textmd{d}\rho$ was found by H\"{u}bner
\cite{hubner}
\begin{equation}\label{hubner}
\textmd{d}_B(\rho,\rho+\textmd{d}\rho)^2=\frac{1}{2}\sum_{i,j=1}^{n}\frac{|\langle
i|\textmd{d}\rho|j\rangle|^2}{\lambda_i+\lambda_j},
\end{equation}
where $\lambda_j$ and $|j\rangle$, $(j=1,2,\cdots,n)$ represent
eigenvalues and eigenvectors of $\rho$, respectively. Dittmann has
derived several explicit formulas, that do not require any
diagonalization procedure, for Bure metric on the manifold of
finite-dimensional nonsingular density matrices
\cite{ditt1,ditt2}, for instance
\begin{equation}\label{ditt2}
\textmd{d}_{\textmd{B}}(\rho,\rho+d\rho)=
\frac{1}{4}\textmd{Tr}\left[\textmd{d}\rho\textmd{d}\rho+
\frac{1}{|\rho|}(\textmd{d}\rho-\rho\textmd{d}\rho)(\textmd{d}\rho-\rho\textmd{d}\rho)\right],
\end{equation}
and
\begin{equation}\label{ditt3}
\begin{array}{rl}
\textmd{d}_{\textmd{B}}(\rho,\rho+d\rho) & =
\frac{1}{4}\textmd{Tr}\left[\textmd{d}\rho\textmd{d}\rho+
\frac{3}{(1-\textmd{Tr}\rho^3)}
(\textmd{d}\rho-\rho\textmd{d}\rho)(\textmd{d}\rho-\rho\textmd{d}\rho)\right.
\\ & + \left.\frac{3|\rho|}{(1-\textmd{Tr}\rho^3)}
(\textmd{d}\rho-\rho^{-1}\textmd{d}\rho)(\textmd{d}\rho-\rho^{-1}\textmd{d}\rho)\right],
\end{array}
\end{equation}
for nonsingular $2\times 2$ and $3\times 3$ density matrices,
respectively.

The probability measure induced by the Bures metric in the space
of mixed quantum states  has been defined by Hall \cite{hall}.
The question of how many entangled or separable states there are
in the set of all quantum ststes is considered by \.Zyczkowski et
al in \cite{karol2,karol3}.  Sommers et al \cite{sommers} have
computed the volume of the $(n^2-1)$-dimensional convex set and
$(n^2-2)$-dimensional hyperarea of the density matrices of an
$n$-level quantum system. In a considerable work, Slater
investigated the use of the volume elements of the Bures metric
as a natural measure over the $(n^2-1)$-dimensional convex set of
$n$-level density matrices, to determine or estimate the volume of
separable states of the pairs of qubit-qubit
\cite{slater2,slater3} and qubit-qutrit \cite{slater4,slater5}.
Very recently \cite{slater6} , Slater made use of the Bloore
parameterization \cite{bloore} of density matrices in order to
obtain the Hilbert-Schmidt volumes of separable subsets for the
two qubit system.

The state space of an $n-$level quantum system is identified with
the set of all $n\times n$ Hermitian positive semidefinite
complex matrices of trace unity, and comprise
$(n^2-1)$-dimensional convex set. Due to considerable interest in
the use of density matrices, a lot of work has been devoted to
describe and parameterize density matrices. Any density matrix of
an n-level system can be expanded in terms of orthogonal
generators $\lambda_i$ of $SU(n)$ as \cite{fano}
\begin{equation}
\rho=\frac{1}{n}\left(I_n+\frac{n(n-1)}{2}\overrightarrow{\nu}\cdot\overrightarrow{\lambda}\right)
\end{equation}
where $\overrightarrow{\nu}=(\nu_1,\nu_2\cdots,\nu_{n^2-1})$ is a
real vector and Lie algebra generators $\lambda_i$ are normalized
as $\textmd{Tr}(\lambda_i\lambda_j)=2\delta_{ij}$. The above
representation is the generalization of the Bloch or coherence
vector representation for two-level systems and gives one of the
possible descriptions of a state on the basis of the actual
measurements which is an important task both from experimental
and theoretical viewpoint \cite{kimura1}. The region of Bloch
vector $\overrightarrow{\nu}$ which represents a physical density
matrix have been found in \cite{kimura1,byrd3}. An investigation
of the geometrical aspects of the Bloch vector space from the
spherical coordinate point of view  is also made by Kimura et al
in \cite{kimura2}.

Boya et al \cite{boya} have shown that the mixed state density
matrices for $n$-level systems can be parameterized in terms of
squared components of an $(n-1)$-sphere and unitary matrices. By
using the Euler angle parameterization of $SU(3)$ group
\cite{byrd1},
\begin{equation}\label{UEuler}
U=\textmd{e}^{i\lambda_3\alpha}\textmd{e}^{i\lambda_2\beta}\textmd{e}^{i\lambda_3\gamma}
\textmd{e}^{i\lambda_5\theta}\textmd{e}^{i\lambda_3
a}\textmd{e}^{i\lambda_2 b} \textmd{e}^{i\lambda_3
c}\textmd{e}^{i\lambda_8\phi/\sqrt{3}},
\end{equation}
where $\lambda_i$ are Gell-Mann matrices,  Byrd et al
\cite{byrd2} have presented a parameterization for
eight-dimensional state space of three-level system as
\begin{equation}\label{RhoEuler}
\rho=V D V^\dag,
\end{equation}
where $D$ is a diagonal density matrix (with two independent
eigenvalues), and $V\in SU(3)$ is given by
\begin{equation}\label{VEuler}
V=\textmd{e}^{i\lambda_3\alpha}\textmd{e}^{i\lambda_2\beta}\textmd{e}^{i\lambda_3\gamma}
\textmd{e}^{i\lambda_5\theta}\textmd{e}^{i\lambda_3
a}\textmd{e}^{i\lambda_2 b}.
\end{equation}
They also gave Bures measure on the space as the product of the
measure on the space of eigenvalues and the truncated Haar
measure on the space of unitary matrices. An Euler angle-based
parameterization for density matrices of four-level system is
also introduced in \cite{tilma1}. A generalized Euler angle
parametrization for $SU(n)$ and $U(n)$ groups has given by Tilma
et al \cite{tilma2,tilma3}. Tilma et al \cite{tilma4} have also
used the parameterization for four-level system (two qubit system)
in order to study the entanglement properties of the system.

In a comprehensive analysis \cite{karol}, \.Zyczkowski et al
analyzed the geometrical properties of the set of mixed quantum
states for an arbitrary $n$-level system and classified the space
of density matrices. Di\c{t}\v{a} \cite{dita1} has provided a
parameterization for general Hermitian operators of n-level
quantum systems. The parameterization is based on the
factorization of $n\times n$ unitary matrices \cite{dita2} and may
be used either for Hamiltonian or density matrices. In
\cite{akhtar}, the authors have shown that the space of two qubit
density matrices (four-level systems), can be characterize with
12-dimensional (as real manifold) space of complex orthogonal
group $SO(4,\mathbb{C})$ together with four positive Wootters's
numbers \cite{woot}, where of course, the normalization condition
reduces the number of parameters to 15.

By using the definitions (\ref{ditt3}) and (\ref{RhoEuler}),
Slater has computed the Bures metric for the eight-dimensional
state space of three-level quantum systems \cite{slater1}. He
showed that all entries of the $8\times 8$ matrix tensor are
independent of the Euler angle $\alpha$, and the matrix tensor
decomposes into a $6\times 6$ block and a $2\times 2$ one, in
corresponding to the six Euler angles of unitary matrix $V$  and
the two independent eigenvalues of diagonal matrix $D$.

In this paper we consider a canonical coset parameterization for
density matrices of an $n$-level quantum system. The
parameterization is based on the coset space decomposition of
unitary matrices \cite{dita1}. This parameterization, as well as
the Euler angle parameterization do, eliminates any
over-parameterization of the density matrix. It also provides a
factorization of the Bures measure on the space of density
matrices as the product of the measure on the space of
eigenvalues and the truncated Haar measure on the space of
unitary matrices. We give explicitly, the parameterization for
two- and three-level density matrices, and by application of
Dittmann's formulas, the Bures metric over the spaces of these
quantum systems are explicitly computed. It is shown that the
coset parameterization gives a  compact expression for the all
metric elements. The analytical simple expression obtained for
the Bures metric of the three-level system enable us to use this
parameterization for the problem of Bures metric over the space
of two qubit system.

The paper is organized as follows: In section 2, the coset space
parameterization of an $n$-level density matrix is introduced. We
give also, explicitly, the parameterization of the density
matrices of the two- and three-level systems in this section.  In
section 3 we compute explicitly the Bures metric of the two- and
the three-level systems. The paper is concluded in section 4 with
a brief conclusion.

\section{Canonical coset parameterization of density matrices}
In this section we review some properties of the set of density
matrices of an $n$-level quantum system and, by using a canonical
coset parameterization for $n\times n$ unitary matrices, we
present a coset parameterization for $n$-level density matrices.
The state space of an $n-$level quantum system is identified with
the set of all $n\times n$ Hermitian positive semidefinite complex
matrices of trace unity, and comprise $(n^2-1)$-dimensional convex
set ${\mathcal M}_n$. The total number of independent variables
needed to parameterize a density matrix $\rho$ is equal to
$n^2-1$, provided no degeneracy occurs.

Let us denote the set of all diagonal density matrices of an
$n$-level system with ${\mathcal D}_n$. An arbitrary element
${D\in {\mathcal D}_n}$ can be written as
\begin{equation}
D=\textmd{diag}\{\lambda_1,\lambda_2,\cdots,\lambda_n\},\qquad
0\le\lambda_i\le 1, \qquad \sum_{i=1}^{n}\lambda_i=1.
\end{equation}
This means that the set of all diagonal density matrices forms an
$(n-1)$-dimensional simplex ${\mathcal S}_{n-1}$ . A generic
density matrix in an arbitrary basis can be obtained as the orbit
of points $D\in {\mathcal D}_n$ under the action of the unitary
group $U(n)$ as
\begin{equation}
\rho=U{D} U^{\dag}.
\end{equation}
Let $H$ be a maximum stability subgroup, i.e. a subgroup of
$U(n)$ that consists of all the group elements $h$ that will
leave the diagonal state $D$ invariant,
\begin{equation}
hDh^\dag =D, \qquad h\in H, \qquad D\in {\mathcal D}_n,
\end{equation}
that is, $H$ contains all elements of $U(n)$ that commute with
$D$. For every element $U\in U(n)$, there is a unique
decomposition of $U$ into a product of two group elements, one in
$H$ and the other in the quotient $G/H$ \cite{gilmore}, i.e.
\begin{equation}
U=\Omega \; h, \qquad U\in U(n), \qquad h\in H, \qquad \Omega\in
U(n)/H.
\end{equation}
The above decomposition implies that the action of an arbitrary
group element $U\in U(n)$ on the point $D\in {\mathcal D}_n$ is
given by
\begin{equation}
\rho=UDU^\dag=\Omega h D h^\dag \Omega^\dag=\Omega D \Omega^\dag.
\end{equation}
This means that in order to characterize the space ${\mathcal
M}_n$, it is sufficient to consider the orbit of points $D\in
{\mathcal D}_n$ under the action of the quotient $U(n)/H$. Since
${\mathcal D}_n$ consists points with different degree of
degeneracy, the maximum stability subgroup will differ for
different $D\in {\mathcal D}_n$ \cite{karol}. Let $m_i$ denotes
degree of degeneracy of eigenvalue $\lambda_i$ of matrix $D$.
This kind of the spectrum follows that $D$ remains invariant
under the action of arbitrary unitary transformation performed in
each of the $m_i$-dimensional eigensubspace. Therefore
$H=U(m_1)\otimes U(m_2)\otimes \cdots U(m_k)$ is maximum
stability subgroup for $D$, and the quotient space $U(n)/H$ is a
complex flag manifold
\begin{equation}
{\mathcal F}=\frac{U(n)}{U(m_1)\otimes U(m_2)\otimes \cdots
U(m_k)},\qquad m_1+m_2+\cdots + m_k=n.
\end{equation}
Two special kinds for the degeneracy of the spectrum of $D$ are as
follows: i) Let $D$ represents the maximally mixed state
$\rho_\ast=\textmd{diag}\{\frac{1}{n},\frac{1}{n},\cdots,\frac{1}{n}\}$.
In this case the stability subgroup $H$ is $U(n)$, and the orbit
of point $\rho_\ast$ is only one point, i.e. $\rho=\rho_\ast$.
ii) On the other hand if the spectrum of $D$ is non-degenerate,
then the stability subgroup is $n$-dimensional torus
$T^n=U(1)^{\otimes n}$, and the orbit of the point $D$ is
\begin{equation}
\rho=\Omega D \Omega^\dag, \qquad \Omega\in U(n)/T^n .
\end{equation}
The maximal torus  $T^n$ is itself a subgroup of all maximum
stability subgroups, therefore the orbit of points $D\in
{\mathcal D}_n$ under the action of quotient $U(n)/T^n$ generates
all points of the space ${\mathcal M}_n$.  The diagonal matrix
$D$ is defined up to a permutation of its entries and, one can
divide the simplex ${\mathcal S}_{n-1}$ into $n!$ identical
simplexes and take any of them. Each part identify points of
${\mathcal S}_{n-1}$ which have the same coordinates, but with
different ordering, and can be considered as the homomorphic
image of simplex ${\mathcal S}_{n-1}$ relative to the discreet
permutation group $P_n$, i.e. ${\mathcal S}_{n-1}/P_n$. Therefore
the points of ${\mathcal M}_n$ can be characterize as the orbit
of diagonal matrices $D\in {\mathcal S}_{n-1}/P_n$ under the
action of quotient $U(n)/T^n$.

Further insight into the space of density matrices can be obtained
by writing  to the arbitrary element $U\in U(n)$ as \cite{gilmore}
\begin{equation}
U=\Omega_n^{(n)} \Omega_{n-1}^{(n)}\cdots
\Omega_2^{(n)}\Omega_1^{(n)},
\end{equation}
where $\Omega_1^{(n)}\in T^n$, and
\begin{equation}
\Omega_{k}^{(n)}\in \frac{U(k)\otimes T^{n-k}}{U(k-1)\otimes
T^{n-k+1}}, \qquad k=2,\cdots,n.
\end{equation}
Comparing this with the decomposition $U=\Omega h$, where $h\in
T^n$, leads to the following decomposition for an arbitrary
element $\Omega$ of quotient $U(n)/T^n$
\begin{equation}
\Omega=\Omega_n^{(n)} \Omega_{n-1}^{(n)}\cdots \Omega_2^{(n)}.
\end{equation}
A typical coset representative $\Omega_k^{(n)}$ can be written as
\cite{gilmore}
\begin{equation}
\Omega_{k}^{(n)}= \left(
\begin{array}{c|c}
SU(k)/U(k-1) & O \\  \hline  O^T & I_{n-k}
\end{array}
\right),
\end{equation}
where $O$, $O^T$ and $I_{n-k}$ represent, respectively, the
$k\times (n-k)$ zero matrix, its transpose and the $(n-k)\times
(n-k)$ identity matrix. The $2(k-1)$-dimensional coset space
$SU(k)/U(k-1)$ have the following $k\times k$ matrix
representation \cite{gilmore}
\begin{equation}\label{cosetB}
SU(k)/U(k-1)= \left(
\begin{array}{c|c}
\cos{\sqrt{B^{(k)} [B^{(k)}]^\dag}} &
B^{(k)}\frac{\sin{\sqrt{[B^{(k)}]^{\dag}
B^{(k)}}}}{\sqrt{[B^{(k)}]^{\dag}B^{(k)}}} \\  \hline
-\frac{\sin{\sqrt{[B^{(k)}]^{\dag}
B^{(k)}}}}{\sqrt{[B^{(k)}]^{\dag}B^{(k)}}}[B^{(k)}]^\dag
&\cos{\sqrt{[B^{(k)}]^{\dag} B^{(k)}}} \\
\end{array}
\right),
\end{equation}
where $B^{(k)}$ represents $(k-1)\times 1$ complex matrix and
$[B^{(k)}]^\dag$ is its adjoint. In the following we consider the
$n=2,3$ cases, explicitly.

\subsection{Two-level system}
We begin by giving the coset parameterization for a two-level
quantum system.  Let us consider a diagonal two-level density
matrix $D=\textmd{diag}\{\lambda_1,\lambda_2\}$. Every $h\in
H=T^2$ leaves the density matrix $D$ invariant i.e. $hDh^\dag=D$
for $h\in H$. Any group element $U\in U(2)$ can be decomposed,
uniquely, as $U=\Omega h$ where $\Omega \in U(2)/T^2$ and $h\in
T^2$ \cite{gilmore}. The coset space with respect to the
stability subgroup $H=T^2$ will provides the unitary
transformations to construct a generic density matrix $\rho$ as
the orbit of diagonal matrix $D$, i.e.
\begin{equation}\label{Rho2x2}
\rho=\Omega D\Omega^\dag, \qquad \Omega\in U(2)/T^2.
\end{equation}
A typical coset
representative in the coset space $U(2)/T^2$ is
\begin{equation}\label{Omega2x2}
\Omega=\left(\begin{array}{cc} \cos{\alpha} &
\textmd{e}^{i\phi}\sin{\alpha} \\
-\textmd{e}^{-i\phi}\sin{\alpha}  & \cos{\alpha}
\end{array}\right),
\end{equation}
where $\alpha, \phi$ are real.  The range of parameters
$\lambda_1,\; \lambda_2$ can be determined as follows. The set of
all diagonal $2\times 2$ matrices $D$ forms a 1-dimensional
simplex ${\mathcal S}_1$, which can be divided into two identical
parts $0\le \lambda_1 \le \frac{1}{2}$, $\frac{1}{2}\le \lambda_1
\le 1$ (see figure 1a). It can be easily seen that each part can
be obtained as the orbit of the other part under the action of the
group element $\Omega=\Omega(\alpha=\phi=\frac{\pi}{2})$ which is
an element of the coset space $U(2)/T^2$. This means that we can
easily consider the diagonal matrix $D$ as
\begin{equation}\label{D2x2}
{D}=\left(\begin{array}{cc}
 \cos^2{\theta} & 0  \\
0 & \sin^2{\theta} \\
\end{array}\right), \qquad 0 \le \theta \le \frac{\pi}{4}.
\end{equation}
With this parameterization any $2\times 2$ density matrix can be
written explicitly  as
\begin{equation}
\rho=\left(\begin{array}{cc}
\sin^2{\alpha}\sin^2{\theta}+\cos^2{\alpha}\cos^2{\theta} &
-\frac{1}{2}\textmd{e}^{i\phi}\sin{2\alpha}\cos{2\theta} \\
-\frac{1}{2}\textmd{e}^{-i\phi}\sin{2\alpha}\cos{2\theta} &
\sin^2{\alpha}\cos^2{\theta}+\cos^2{\alpha}\sin^2{\theta}
\end{array}
\right).
\end{equation}

\subsection{Three-level system}
The density matrices for the three-level system comprise
eight-dimensional convex set. Let
$D=\textmd{diag}\{\lambda_1,\lambda_2,\lambda_3\}$ be a $3\times
3$ diagonal density matrix of a three-level system. The density
matrix $D$ is invariant under the action of every group element
$h\in H=T^3$, i.e. $hDh^\dag=D$ for $h\in H$. The coset
decomposition $U=\Omega h$ where $\Omega \in U(3)/T^3$, $h\in
T^3$ provides a parameterization for a generic $\rho$ as
\begin{equation}\label{Rho3x3}
\rho=U D U^{\dag}=\Omega h D h^\dag \Omega^\dag= \Omega D
\Omega^\dag.
\end{equation}
On the other hand the decomposition
\begin{equation}
U=\Omega_3^{(3)}\Omega_2^{(3)}\Omega_1^{(3)},
\end{equation}
with
\begin{equation}
\begin{array}{l}
\Omega_3^{(3)}\in U(3)/U(2)\otimes U(1), \\
\Omega_2^{(3)}\in (U(2)\otimes U(1))/(U(1)\otimes U(1)\otimes U(1)), \\
\Omega_1^{(3)}\in U(1)\otimes U(1)\otimes U(1),
\end{array}
\end{equation}
follows that
\begin{equation}\label{Omega3x3}
\Omega=\Omega_3^{(3)}\Omega_2^{(3)}.
\end{equation}
The coset representatives $\Omega_2^{(3)}$ and $\Omega_3^{(3)}$
can be parameterized respectively as
\begin{equation}
\Omega_2^{(3)}=\left(\begin{array}{ccc}
 \cos{\alpha} &
e^{i\phi}\sin{\alpha} & 0 \\
-e^{-i\phi}\sin{\alpha} & \cos{\alpha} & 0 \\
0 & 0 & 1
\end{array}\right),
\end{equation}
and
\begin{equation}
\Omega_3^{(3)}=\left(\begin{array}{ccc}
 1+\frac{\beta_1^2}{\beta^2}(\cos{\beta}-1) &
\frac{\beta_1\beta_2}{\beta^2}e^{i(\psi_1-\psi_2)}(\cos{\beta}-1)
&
\frac{\beta_1}{\beta}e^{i\psi_1}\sin{\beta} \\
\frac{\beta_1\beta_2}{\beta^2}e^{-i(\psi_1-\psi_2)}(\cos{\beta}-1)
& 1+\frac{\beta_2^2}{\beta^2}(\cos{\beta}-1) &
\frac{\beta_2}{\beta}e^{i\psi_2}\sin{\beta} \\
-\frac{\beta_1}{\beta}e^{-i\psi_1}\sin{\beta} &
-\frac{\beta_2}{\beta}e^{-i\psi_2}\sin{\beta} & \cos{\beta}
\end{array}\right),
\end{equation}
where $\beta=\sqrt{\beta_1^2+\beta_2^2}$.

The two-dimensional simplex  ${\mathcal S}_2$ of eigenvalues of
$\rho$ is divided into $3!$ parts (see figure 1b).  Let us take
one of the parts (e.g. the shaded one) for illustration. It is
easy to see that all other parts can be obtained from this one by
applying the permutation group $P_3$. On the other hand the
elements of the permutation group $P_3$ can be obtained from the
coset representative (\ref{Omega3x3}) up to a phase  as
$$
\begin{array}{l}
\Omega_3^{(3)}(\beta_1=\beta_2=0)\Omega_2^{(3)}(\alpha=0)=(Id),
\\
\Omega_3^{(3)}(\beta_1=\beta_2=0)\Omega_2^{(3)}(\alpha=\phi=\frac{\pi}{2})=i(12),
\\
\Omega_3^{(3)}(\beta_1=\frac{\pi}{2},\beta_2=0,\psi_1=\frac{\pi}{2})
\Omega_2^{(3)}(\alpha=0)=i(13),
\\
\Omega_3^{(3)}(\beta_1=0,\beta_2=\frac{\pi}{2},\psi_2=\frac{\pi}{2})
\Omega_2^{(3)}(\alpha=0)=i(23),
\\
\Omega_3^{(3)}(\beta_1=\frac{\pi}{2},\beta_2=0,\psi_1=\frac{\pi}{2})
\Omega_2^{(3)}(\alpha=\phi=\frac{\pi}{2})=i(123),
\\
\Omega_3^{(3)}(\beta_1=0,\beta_2=\frac{\pi}{2},\psi_2=\frac{\pi}{2})
\Omega_2^{(3)}(\alpha=\phi=\frac{\pi}{2})=i(321).
\end{array}
$$
Therefore the ranges of the eigenvalues of $D$ can be determined
as $\frac{1}{3}\le \lambda_1\le 1$, $0\le \lambda_2\le
\frac{1}{2}$ and $0\le \lambda_3\le \frac{1}{3}$, or
equivalently, one can parameterize  the diagonal matrix $D$ as
\begin{equation}\label{D3x3}
{D}=\left(\begin{array}{ccc}
\cos^2{\theta_1} & 0 & 0 \\
0 & \sin^2{\theta_1}\cos^2{\theta_2} & 0 \\
0 & 0 & \sin^2{\theta_1}\sin^2{\theta_2}
\end{array}\right),
\end{equation}
where $0\le \theta_1\le \cos^{-1}{\frac{1}{\sqrt{3}}},\;\;
\frac{\pi}{6}\le \theta_2 \le \frac{\pi}{4}$.

\section{Bures metric}
In this section we calculate the Bures metric of the two- and the
three-level quantum systems.  We will use the canonical coset
parameterization of the density matrices introduced in the last
section.
\subsection{Two-level system}
In the two-level systems by using Eqs. (\ref{ditt2}) and
(\ref{Rho2x2}) we get
\begin{equation}
\begin{array}{rl}
\textmd{d}_{\textmd{B}}(\rho,\rho+d\rho) & =
\frac{1}{4}\textmd{Tr}\left[(\textmd{d}{D})^2
+\frac{1}{|{D}|}({\mathcal I}-{D})^2(\textmd{d}{D})^2\right] \\
& +\sum_{i<j}\left[S_{ij}\left((\Omega^\dag\textmd{d}\Omega)_{ij}
(\Omega^\dag\textmd{d}\Omega)_{ji}\right)\right],
\end{array}
\end{equation}
where ${\mathcal I}$ is unit matrix and
\begin{equation}
S_{ij}=
\frac{1}{|{D}|}\left[({D}_{ii}-{D}_{jj})^2({D}_{ii}+{D}_{jj}-{D}_{ii}{D}_{jj}-|{D}|-1)\right].
\end{equation}
By defining the ordering $\{\theta, \alpha, \phi\}$ for
coordinates, the corresponding Bures metric tensor takes the
following diagonal form
\begin{equation}
g=\left(\begin{array}{ccc} 1 & 0 & 0 \\
0 & \cos^2{2\theta} & 0 \\
0 & 0 & \frac{1}{4}\sin^2{2\alpha}\cos^2{2\theta}
\end{array}\right).
\end{equation}
\begin{figure}[h]
\vspace{-70mm}
\centerline{\includegraphics[height=26cm]{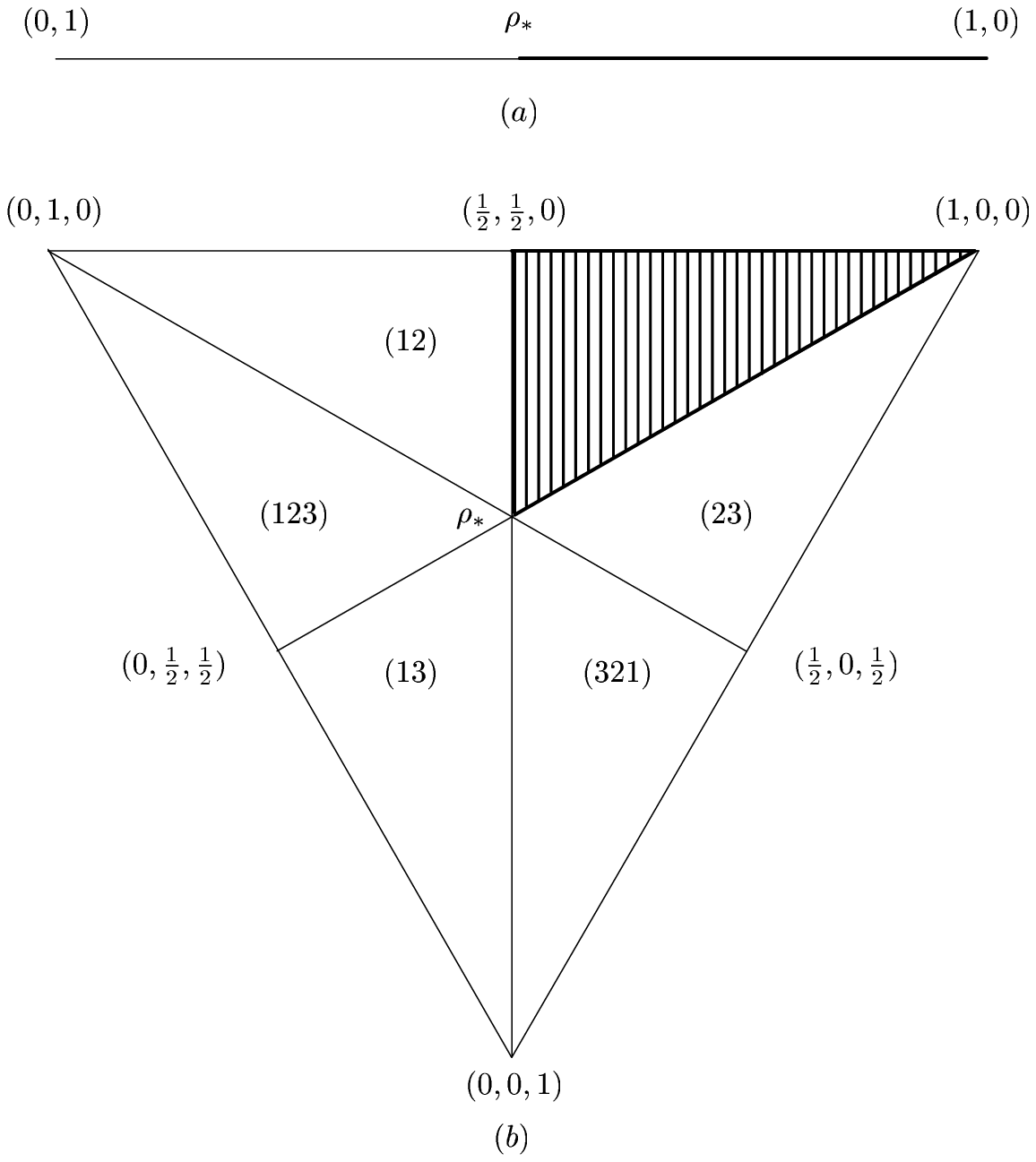}}
\vspace{-70mm} \caption{(a) One-dimensional simplex ${\mathcal
S}_1$ of diagonal density matrices of two-level systems, (b)  the
two-dimensional simplex ${\mathcal S}_2$ of diagonal density
matrices of three-level systems. The simplex ${\mathcal S}_n$ can
be decomposed into $n!$ parts. The parts can be transformed to
each other by applying the elements of the permutation group
$P_n$.}
\end{figure}

\subsection{Three-level system}
In the three-level system, by using (\ref{Rho3x3}) in
(\ref{ditt3}) we have
\begin{equation}
\begin{array}{rl}
\textmd{d}_{\textmd{B}}(\rho,\rho+d\rho) & =
\frac{1}{4}\textmd{Tr}\left[(\textmd{d}{D})^2
+\frac{3}{(1-\textmd{Tr}{D}^3)}\left(({\mathcal
I}-{D})^2(\textmd{d}{D})^2
+|{D}|({\mathcal I}-{D}^{-1})^2(\textmd{d}{D})^2\right)\right] \\
& +\sum_{i<j}\left[T_{ij}
\left((\Omega^\dag\textmd{d}\Omega)_{ij}(\Omega^\dag\textmd{d}\Omega)_{ji}\right)\right],
\end{array}
\end{equation}
where
\begin{equation}
\begin{array}{rl}
T_{ij} =\frac{3}{2(1-\textmd{Tr}{D}^3)} & \left[
({D}_{ii}-{D}_{jj})^2({D}_{ii}+{D}_{jj}-{D}_{ii}{D}_{jj}-\textmd{Tr}{D}^3-2)\right.
\\
& +\left.|{D}|({D}^{-1}_{ii}-{D}^{-1}_{jj})^2({D}^{-1}_{ii}
+{D}^{-1}_{jj}-{D}^{-1}_{ii}{D}^{-1}_{jj}-1)\right].
\end{array}
\end{equation}
Now defining  the ordering of coordinates as $\{\theta_1,
\theta_2, \alpha, \phi, \beta_1, \beta_2, \psi_1, \psi_2\} $, the
corresponding symmetric matrix for metric tensor takes the
following form
\begin{equation} g=\left(
\begin{array}{cc|cccccc}
1 & 0 & 0 & 0 & 0 & 0 & 0 & 0 \\
 & \sin^2{\theta_1} & 0 & 0 & 0 & 0 & 0 & 0 \\ \hline
  &  & g_{\alpha\alpha} & 0 & g_{\alpha\beta_1} &
  g_{\alpha\beta_2} & g_{\alpha\psi_1} & g_{\alpha\psi_2} \\
&  &  & g_{\phi\phi} & g_{\phi\beta_1} &
  g_{\phi\beta_2} & g_{\phi\psi_1} & g_{\phi\psi_2} \\
  &  &  &  & g_{\beta_1\beta_1} &
  g_{\beta_1\beta_2} & g_{\beta_1\psi_1} & g_{\beta_1\psi_2} \\
  &  &  &  &  &
  g_{\beta_2\beta_2} & g_{\beta_2\psi_1} & g_{\beta_2\psi_2} \\
  &  &  &  &  &
   & g_{\psi_1\psi_1} & g_{\psi_1\psi_2} \\
    &  &  &  &  &
   &  & g_{\psi_2\psi_2}
\end{array}  \right).
\end{equation}
The $8\times 8$ matrix tensor decomposes into a $2\times 2$ block
and a $6\times 6$ one, in corresponding to the two independent
eigenvalues $\theta_1,\theta_2$ and six coset parameters $\alpha,
\phi, \beta_1, \beta_2, \psi_1, \psi_2$. After some analytical
calculations, we can obtain the following expression for  the
matrix elements of the $6\times 6$ block
\begin{itemize}
\item
$g_{\alpha\alpha}=-T_{12}$

\item
$g_{\alpha\phi}=0$

\item
$g_{\alpha\beta_1}=
2T_{12}\beta_2\cos{\gamma}\left(\frac{\sin{\frac{\beta}{2}}}{\beta}\right)^2$

\item
$g_{\alpha\beta_2}=-\frac{\beta_1}{\beta_2}g_{\alpha\beta_1}$

\item
$g_{\alpha\psi_1}=
2T_{12}\beta_1\beta_2U_2\sin{\gamma}\left(\frac{\sin{\frac{\beta}{2}}}{\beta}\right)^2$

\item
$g_{\alpha\psi_2}=\frac{U_1}{U_2}g_{\alpha\psi_1}$

\item
$g_{\phi\phi}=\frac{-1}{4}T_{12}\sin^2{2\alpha}$

\item
$g_{\phi\beta_1}=
\frac{-1}{2}T_{12}\beta_2\sin{4\alpha}\left(\frac{\sin{\frac{\beta}{2}}}{\beta}\right)^2$

\item
$g_{\phi\beta_2}=-\frac{\beta_1}{\beta_2}g_{\phi\beta_1}$

\item
$g_{\phi\psi_1}= \frac{1}{2}T_{12}\beta_1\sin{2\alpha}
\left[\beta_1W_2\sin{2\alpha}+2\beta_2U_2\cos{2\alpha}\cos{\gamma}\right]
\left(\frac{\sin{\frac{\beta}{2}}}{\beta}\right)^2$

\item
$g_{\phi\psi_2}= \frac{-1}{2}T_{12}\beta_2\sin{2\alpha}
\left[\beta_2W_1\sin{2\alpha}-2\beta_1U_1\cos{2\alpha}\cos{\gamma}\right]
\left(\frac{\sin{\frac{\beta}{2}}}{\beta}\right)^2$

\item
$g_{\beta_1\beta_1}  =  -4T_{12}\beta_2^2
\left[1-\sin^2{2\alpha}\sin^2{\gamma}\right]
\left(\frac{\sin{\frac{\beta}{2}}}{\beta}\right)^4$
$$\hspace{-47mm}
\begin{array}{rl}
 &
-T_{13}\left[X^2\sin^2{\alpha}+V_1^2\cos^2{\alpha}-XV_1\sin{2\alpha}\cos{\gamma}\right]
\\
& -T_{23}
\left[X^2\cos^2{\alpha}+V_1^2\sin^2{\alpha}+XV_1\sin{2\alpha}\cos{\gamma}\right]
\end{array}
$$

\item
$g_{\beta_1\beta_2}
=4T_{12}\beta_1\beta_2\left[1-\sin^2{2\alpha}\sin^2{\gamma}\right]
\left(\frac{\sin{\frac{\beta}{2}}}{\beta}\right)^4$
$$\hspace{-28mm}
\begin{array}{rl}
 & -T_{13}
\left[X(V_1\cos^2{\alpha}+V_2\sin^2{\alpha})
-\frac{1}{2}(V_1V_2+X^2)\sin{2\alpha}\cos{\gamma}\right] \\
 & -T_{23}
\left[X(V_1\sin^2{\alpha}+V_2\cos^2{\alpha})
+\frac{1}{2}(V_1V_2+X^2)\sin{2\alpha}\cos{\gamma}\right]
\end{array}
$$

\item
$g_{\beta_1\psi_1}=-T_{12}\beta_1\beta_2
\left[2\beta_2U_2\sin^2{2\alpha}\sin{2\gamma}-\beta_1W_2\sin{4\alpha}\sin{\gamma}\right]
\left(\frac{\sin{\frac{\beta}{2}}}{\beta}\right)^4$
$$\hspace{-41mm}
\begin{array}{rl}
& +\frac{1}{2}(T_{13}-T_{23})\beta_1\sin{2\alpha}\sin{\gamma}
\left[U_2X+V_1Y\right]\left(\frac{\sin{\beta}}{\beta}\right)
\end{array}
$$

\item
$g_{\beta_1\psi_2}  =-T_{12}\beta_2^2
\left[2\beta_1U_1\sin^2{2\alpha}\sin{2\gamma}+\beta_2W_1\sin{4\alpha}\sin{\gamma}\right]
\left(\frac{\sin{\frac{\beta}{2}}}{\beta}\right)^4$
$$\hspace{-45mm}
\begin{array}{rl}
& -\frac{1}{2}(T_{13}-T_{23})\beta_2\sin{2\alpha}\sin{\gamma}
\left[U_1V_1+XY\right]\left(\frac{\sin{\beta}}{\beta}\right)
\end{array}
$$

\item
$g_{\beta_2\beta_2} = -4T_{12}\beta_1^2
\left[1-\sin^2{2\alpha}\sin^2{\gamma}\right]
\left(\frac{\sin{\frac{\beta}{2}}}{\beta}\right)^4$
$$\hspace{-47mm}
\begin{array}{rl}
 &
-T_{13}\left[X^2\cos^2{\alpha}+V_2^2\sin^2{\alpha}-XV_2\sin{2\alpha}\cos{\gamma}\right]
\\
&-T_{23}
\left[X^2\sin^2{\alpha}+V_2^2\cos^2{\alpha}+XV_2\sin{2\alpha}\cos{\gamma}\right]
\end{array}
$$

\item
$g_{\beta_2\psi_1} =T_{12} \beta_1^2
\left[2\beta_2U_2\sin^2{2\alpha}\sin{2\gamma}-\beta_1W_2\sin{4\alpha}\sin{\gamma}\right]
\left(\frac{\sin{\frac{\beta}{2}}}{\beta}\right)^4$
$$\hspace{-44mm}
\begin{array}{rl}
&+\frac{1}{2}(T_{13}-T_{23})\beta_1\sin{2\alpha}\sin{\gamma}
\left[U_2V_2+XY\right]\left(\frac{\sin{\beta}}{\beta}\right)
\end{array}
$$

\item
$g_{\beta_2\psi_2} =T_{12}\beta_1\beta_2
\left[2\beta_1U_1\sin^2{2\alpha}\sin{2\gamma}+\beta_2W_1\sin{4\alpha}\sin{\gamma}\right]
\left(\frac{\sin{\frac{\beta}{2}}}{\beta}\right)^4$
$$\hspace{-45mm}
\begin{array}{rl}
&-\frac{1}{2}(T_{13}-T_{23})\beta_2\sin{2\alpha}\sin{\gamma}
\left[U_1X+V_2Y\right] \left(\frac{\sin{\beta}}{\beta}\right)\\
\end{array}
$$

\item
$g_{\psi_1\psi_1}=  -T_{12}\beta_1^2
\left[4\beta_2^2U_2^2(1-\sin^2{2\alpha}\cos^2{\gamma})\right.$
$$\hspace{-31mm}
\begin{array}{rl}
 & \left.+\beta_1^2W_2^2\sin^2{2\alpha}
+2\beta_1\beta_2U_2W_2\sin{4\alpha}\cos{\gamma}\right]
\left(\frac{\sin{\frac{\beta}{2}}}{\beta}\right)^4 \\
&-T_{13}\beta_1^2
\left[U_2^2\cos^2{\alpha}+Y^2\sin^2{\alpha}+U_2Y\sin{2\alpha}\cos{\gamma}\right]
\left(\frac{\sin{\beta}}{\beta}\right)^2
\\
&-T_{23}\beta_1^2
\left[U_2^2\sin^2{\alpha}+Y^2\cos^2{\alpha}-U_2Y\sin{2\alpha}\cos{\gamma}\right]
\left(\frac{\sin{\beta}}{\beta}\right)^2
\end{array}
$$

\item
$g_{\psi_1\psi_2}= -T_{12}\beta_1\beta_2
\left[4\beta_1\beta_2U_1U_2(1-\sin^2{2\alpha}\cos^2{\gamma})\right.$
$$\hspace{-8mm}
\begin{array}{rl}
& \left.-\beta_1\beta_2W_1W_2\sin^2{2\alpha}
-\sin{4\alpha}\cos{\gamma}(\beta_2^2U_2W_1-\beta_1^2U_1W_2)\right]
\left(\frac{\sin{\frac{\beta}{2}}}{\beta}\right)^4
\\
&+T_{13}\beta_1\beta_2
\left[Y(U_1\sin^2{\alpha}+U_2\cos^2{\alpha})
+\frac{1}{2}\sin{2\alpha}\cos{\gamma}(U_1U_2+Y^2)\right]
\left(\frac{\sin{\beta}}{\beta}\right)^2
\\
&+T_{23}\beta_1\beta_2
\left[Y(U_1\cos^2{\alpha}+U_2\sin^2{\alpha})
-\frac{1}{2}\sin{2\alpha}\cos{\gamma}(U_1U_2+Y^2)\right]
\left(\frac{\sin{\beta}}{\beta}\right)^2
\end{array}
$$

\item
$g_{\psi_2\psi_2}= -T_{12}\beta_2^2
\left[4\beta_1^2U_1^2(1-\sin^2{2\alpha}\cos^2{\gamma})\right.$
$$\hspace{-30mm}
\begin{array}{rl}
 & \left.+\beta_2^2W_1^2\sin^2{2\alpha}
-2\beta_1\beta_2U_1W_1\sin{4\alpha}\cos{\gamma}\right]
\left(\frac{\sin{\frac{\beta}{2}}}{\beta}\right)^4 \\
&-T_{13}\beta_2^2
\left[U_1^2\sin^2{\alpha}+Y^2\cos^2{\alpha}+U_1Y\sin{2\alpha}\cos{\gamma}\right]
\left(\frac{\sin{\beta}}{\beta}\right)^2
\\
&-T_{23}\beta_2^2
\left[U_1^2\cos^2{\alpha}+Y^2\sin^2{\alpha}-U_1Y\sin{2\alpha}\cos{\gamma}\right]
\left(\frac{\sin{\beta}}{\beta}\right)^2.
\end{array}
$$
\end{itemize}
In the above equations we have used the following definitions
\begin{equation}
\gamma=\phi-\psi_1+\psi_2,
\end{equation}
and
\begin{equation}
\begin{array}{l}
T_{12}=\frac{-1}{2}\left(\cos^2{\theta_1}-\sin^2{\theta_1}\cos^2{\theta_2}\right)^2
\left[3+\frac{(1-\sin^2{\theta_1}\cos^2{\theta_2})
(1+\cos^4{\theta_1}\cos^4{\theta_2})}
{\cos^4{\theta_1}\cos^4{\theta_2}
\left(\cos^2{\theta_1}+\sin^4{\theta_1}\sin^2{\theta_2}\cos^2{\theta_2}\right)
}\right],
\\
T_{13}=\frac{-1}{2}\left(\cos^2{\theta_1}-\sin^2{\theta_1}\sin^2{\theta_2}\right)^2
\left[3+\frac{(1-\sin^2{\theta_1}\sin^2{\theta_2})
(\cos^2{\theta_2}+\sin^2{\theta_1}\cos^4{\theta_1}\sin^4{\theta_2})}
{\sin^2{\theta_1}\cos^4{\theta_1}\cos^4{\theta_2}
\left(\cos^2{\theta_1}+\sin^4{\theta_1}\sin^2{\theta_2}\cos^2{\theta_2}\right)}\right],
\\
T_{23}=\frac{-1}{2}\left[\sin^2{\theta_1}\cos^2{\theta_2}(1+3\sin^2{\theta_1})
+\frac{\cos^2{\theta_1}}
{\sin^6{\theta_1}\sin^4{\theta_2}\cos^2{\theta_2}}\right],
\end{array}
\end{equation}
and
\begin{equation}
\begin{array}{ll}
U_1=\frac{\beta_1^2}{\beta^2}+\frac{\beta_2^2}{\beta^2}\cos{\beta},
&
U_2=\frac{\beta_2^2}{\beta^2}+\frac{\beta_1^2}{\beta^2}\cos{\beta}, \\
V_1=\frac{\beta_1^2}{\beta^2}+\frac{\beta_2^2}{\beta^2}\left(\frac{\sin{\beta}}{\beta}\right),
&
V_2=\frac{\beta_2^2}{\beta^2}+\frac{\beta_1^2}{\beta^2}\left(\frac{\sin{\beta}}{\beta}\right),
\\
W_1=(1+\cos{\beta})+2\frac{\beta_1^2}{\beta^2}(1-\cos{\beta}), &
W_2=(1+\cos{\beta})+2\frac{\beta_2^2}{\beta^2}(1-\cos{\beta}), \\
X=\frac{\beta_1}{\beta}\frac{\beta_2}{\beta}(1-\frac{\sin{\beta}}{\beta}),
& Y=\frac{\beta_1}{\beta}\frac{\beta_2}{\beta}(1-\cos{\beta}).
\end{array}
\end{equation}
It should be stress that all elements $g_{ij}$ are simply products
of two independent functions, one of the coset space parameters,
and the other of the spherical angles $\theta_1$, $\theta_2$. It
is also worth to note that  the three angles $\phi$, $\psi_1$ and
$\psi_2$ appear only in the form $\gamma=\phi-\psi_1+\psi_2$.

\section{Conclusion}
We present a coset parameterization for density matrices of an
$n$-level quantum system. The parameterization is based on the
canonical coset decomposition of unitary matrices. By using the
parameterization for two- and three-level quantum systems, the
Bures metric over the state space of these systems are computed
explicitly. We show that in the canonical coset parameterization
the symbolic expression for all tensor elements can be obtained.
The problem of computing the Bures metric of a two-qubit
(four-level)  system, which is important for calculations
involving entanglement, is also under consideration.


\begin{thebibliography}{99}
\bibitem{bures}{ D. J. C. Bures,}
{\em Trans. Am. Math. Phys. {\bf 135}, 199 (1969).}
\bibitem{braun}{ S. L. Braunstein and C. M. Caves,}
{\em Phys. Rev. Lett.  {\bf 72}, 3439 (1994).}
\bibitem{petz}{ D. Petz and C. Sud\'{a}r,}
{\em J. Math. Phys.  {\bf 37}, 2662 (1996).}
\bibitem{uhlmann}{ A. Uhlmann,}
{\em Rep. Math. Phys.  {\bf 9}, 273 (1976).}
\bibitem{jozsa}{ R. Jozsa,}
{\em J. Mod. Opt.  {\bf 41}, 2315 (1994).}
\bibitem{hubner}{ M. H\"{u}bner,}
{\em Phys. Lett. A  {\bf 163}, 239 (1992).}
\bibitem{ditt1}{ J. Dittmann,}
{\em Sem. Sophus Lie  {\bf 3}, 73 (1993).}
\bibitem{ditt2}{ J. Dittmann,}
{\em J. Phys. A: Math. Gen.  {\bf 32}, 2663 (1999).}
\bibitem{hall}{ M. J. W. Hall,}
{\em Phys. Lett. A {\bf 242}, 123 (1998).}
\bibitem{karol2}{ K. \.{Z}yczkowski, P. Horodecki, A. Sanpera and M. Lewenstein,}
{\em J. Phys. Rev. A {\bf 58}, 883 (1998).}
\bibitem{karol3}{ K. \.{Z}yczkowski,}
{\em Phys. Rev. A {\bf 60}, 3496 (1999).}
\bibitem{sommers}{ H-J Sommers and K. \.{Z}yczkowski,}
{\em J. Phys. A: Math. Gen. {\bf 36}, 10083 (2003).}

\bibitem{slater2}{ P. B. Slater,}
{\em Quantum Inf. Process. {\bf 1}, 387 (2002).}
\bibitem{slater3}{ P. B. Slater,}
{\em J. Geom. Phys. {\bf 53}, 74 (2005).}
\bibitem{slater4}{ P. B. Slater,}
{\em J. Opt. B: Quantum Semiclass. Opt. {\bf 5}, S651 (2003).}
\bibitem{slater5}{ P. B. Slater,}
{\em Phys. Rev. A {\bf 71}, 052319 (2005).}
\bibitem{slater6}{ P. B. Slater,}
{\em e-print quant-ph/0609006.}
\bibitem{bloore}{ F. J. Bloore,}
{\em J. Phys. A: Math. Gen. {\bf 9}, 2059 (1976).}



\bibitem{fano}{ U. Fano,}
{\em Rev. Mod. Phys. {\bf 29}, 74 (1957).}
\bibitem{kimura1}{ G. Kimura,}
{\em Phys. Lett. A {\bf 314}, 339 (2003).}
\bibitem{byrd3}{ M. S. Byrd and N. Khaneja,}
{\em Phys. Rev. A {\bf 68}, 062322 (2003).}
\bibitem{kimura2}{ G. Kimura and A. Kossakowski,}
{\em Open Sys. Information Dyn. {\bf 12}, 207 (2005).} {\em J.
Phys. A: Math. Gen.  {\bf 32}, 2663 (1999).}
\bibitem{boya}{ L. J. Boya, M. Byrd, M. Mims and E. C. G. Sudarshan,}
{\em quant-ph/9810084}
\bibitem{byrd1}{ M. S. Byrd,}
{\em J. Math. Phys. {\bf 39}, 6125 (1998), {\it ibid} {\bf 41},
1026 (2000) Erratum.}
\bibitem{byrd2}{ M. S. Byrd and P. Slater,}
{\em Phys. Lett. A  {\bf 283}, 152 (2001).}
\bibitem{tilma1}{ T. Tilma, M. Byrd and E. C. G. Sudarshan,}
{\em J. Phys. A: Math. Gen. {\bf 35}, 10445 (2002).}
\bibitem{tilma2}{ T. Tilma and E. C. G. Sudarshan,}
{\em J. Phys. A: Math. Gen. {\bf 35}, 10467 (2002).}
\bibitem{tilma3}{ T. Tilma and E. C. G. Sudarshan,}
{\em J. Geom. Phys. {\bf 52}, 263 (2004).}
\bibitem{tilma4}{ T. Tilma and E. C. G. Sudarshan,}
{\em J Phys. Soc. Jpn. {\bf 72}, Suppl. C, 185 (2003).}
\bibitem{karol}{ K. \.{Z}yczkowski and W. S\l omczy\'nski,}
{\em J. Phys. A: Math. Gen. {\bf 34}, 6689 (2001).}
\bibitem{dita1}{ P. Di\c{t}\v{a},}
{\em J. Phys. A: Math. Gen. {\bf 38}, 2657 (2005).}
\bibitem{dita2}{ P. Di\c{t}\v{a},}
{\em J. Phys. A: Math. Gen. {\bf 36}, 1 (2003).}
\bibitem{akhtar}{ S. J. Akhtarshenas and M. A. Jafarizadeh,}
{\em Quantum Inf. Comput. {\bf 3}, 229 (2003).}
\bibitem{woot}{ W. K. Wootters, }
{\em Phys. Rev. Lett. {\bf 80} 2245 (1998).}
\bibitem{slater1}{ P. Slater,}
{\em J. Geom. Phys. {\bf 39}, 207 (2001).}
\bibitem{gilmore}{ R. Gilmore,}
{\em ``Lie Groups, Lie Algebras, and Some of Their
Applications''}, John-Wiley Publishing Co., (1974).
\end{thebibliography}
\end{document}